\documentclass[a4paper, 10pt, conference]{IEEEtran}
\IEEEoverridecommandlockouts
\usepackage{cite}
\usepackage{amsmath,amssymb,amsfonts}
\usepackage{algorithmic}
\usepackage{graphicx}
\usepackage{textcomp}
\usepackage{xcolor}
\usepackage{float}
\usepackage{booktabs}
\usepackage{cleveref}
\usepackage{multirow}
\usepackage{subfigure}
\usepackage{balance}
\usepackage{geometry}
\begin{document}
\title{Towards Real-world Deployment of NILM Systems: Challenges and Practices\\
\thanks{Corresponding author: Guoming Tang (guomingtang@hkust-gz.edu.cn)}
}
\author{\IEEEauthorblockN{Junyu Xue\IEEEauthorrefmark{1}\IEEEauthorrefmark{2},
Yu Zhang\IEEEauthorrefmark{4},
Xudong Wang\IEEEauthorrefmark{5}, 
Yi Wang\IEEEauthorrefmark{1}\IEEEauthorrefmark{2}\IEEEauthorrefmark{3},
Guoming Tang\IEEEauthorrefmark{6}}
\IEEEauthorblockA{\IEEEauthorrefmark{1}Southern University of Science and Technology, Shenzhen, China}
\IEEEauthorblockA{\IEEEauthorrefmark{2}Pengcheng Laboratory, Shenzhen, China}
\IEEEauthorblockA{\IEEEauthorrefmark{3}Heyuan Bay Area Digital Economy Technology Innovation Center, Heyuan, China}
\IEEEauthorblockA{\IEEEauthorrefmark{4}The Chinese University of Hong Kong, Hong Kong, China}
\IEEEauthorblockA{\IEEEauthorrefmark{5}The Chinese University of Hong Kong, Shenzhen, China}
\IEEEauthorblockA{\IEEEauthorrefmark{6}The Hong Kong University of Science and Technology (Guangzhou), Guangzhou, China}
}
\maketitle
\begin{abstract}
Non-intrusive load monitoring (NILM), as a key load monitoring technology, can much reduce the deployment cost of traditional power sensors. Previous research has largely focused on developing cloud-exclusive NILM algorithms, which often result in high computation costs and significant service delays. To address these issues, we propose a three-tier framework to enhance the real-world applicability of NILM systems through edge-cloud collaboration. Considering the computational resources available at both the edge and cloud, we implement a lightweight NILM model at the edge and a deep learning based model at the cloud, respectively. In addition to the differential model implementations, we also design a NILM-specific deployment scheme that integrates Gunicorn and NGINX to bridge the gap between theoretical algorithms and practical applications. To verify the effectiveness of the proposed framework, we apply real-world NILM scenario settings and implement the entire process of data acquisition, model training, and system deployment. The results demonstrate that our framework can achieve high decomposition accuracy while significantly reducing the cloud workload and communication overhead under practical considerations.
\end{abstract}
\section{Introduction}
To monitor the status and energy consumption of household electrical appliances, load monitoring technologies are widely used to analyze energy system data. Among these, Non-Intrusive Load Monitoring (NILM) is a crucial research area, as it eliminates the need for separate sensors for each appliance. In general, NILM is highly significant for energy conservation, emission reduction, fault diagnosis, and power grid planning. Particularly, by providing detailed consumption data, NILM can achieve significant energy savings of around 15\%, as it encourages households to use energy more sustainably\cite{b1}. Furthermore, NILM can enhance protection plans, improve load forecasting accuracy, and serve as a benchmark for grid management. With real-time NILM, utilities can recommend specific appliance operations, such as switching off air conditioners during peak hours, to manage the power load more effectively\cite{b2}.

Although the concept of NILM was introduced in the past, its widespread application was initially limited by low accuracy and scarce data resources. However, recent advancements in big data, deep learning, and related technologies have led to an abundance of electrical data and significant improvements in NILM accuracy and decomposition, thus promoting its applications. To further enable the wide application of NILM in reality, distributed learning technologies, particularly edge-cloud collaboration, are becoming essential. Cloud computing provides robust computational power and inferential capabilities for training large machine-learning models used in NILM, allowing enterprises to enhance model accuracy and broaden service scope based on data characteristics and scale. Meanwhile, edge computing preprocesses electrical data at the edge, reducing the load on the cloud. With adequate edge computing power, lightweight models can be deployed locally to minimize data transmission delays and mitigate data leakage risks. This collaboration leverages the strengths of both technologies, offering users more efficient and secure services.

Several studies have combined NILM with edge-cloud collaboration techniques. Hong et al.\cite{b3} proposed a NILM framework based on edge-cloud collaboration, utilizing the deep learning Adaboost algorithm for matching at the edge. However, this framework relies on high-frequency sampling data to improve accuracy, resulting in high communication costs. Hudson et al.\cite{b4} designed a framework that integrates federated learning with edge-cloud collaboration to address NILM challenges, reducing communication costs and privacy leakage risks, but it does not thoroughly discuss the cooperative functioning of edge and cloud components. Generally, current research focuses on cloud-exclusive functions, leading to high consumption of cloud computing resources and significant delays. Moreover, these studies often overlook real-world settings and lack practical applications. Motivated by this, we emphasize the practical significance of edge-cloud collaboration in addressing NILM application challenges, particularly in balancing resource consumption and managing concurrency issues. Our contributions can be summarized as follows:

\begin{itemize}
    \item We propose a three-tier collaborative framework involving the cloud, edge, and client to enhance the applicability of NILM technology, aiming to reduce the cloud's workload and minimize data transmission delays.
    \item We present a real-world deployment scheme to promote NILM in practical applications, integrating Gunicorn and NGINX components to address response time and load balancing issues from the cloud to the edge.
    \item We evaluate the proposed NILM framework on real-world datasets and appliance settings. The results from both models deployed at the cloud and edge demonstrate that: only through edge-cloud collaboration can we achieve both accurate performance and low-latency response. We also find that combining NGINX and Gunicorn can significantly improve concurrency performance compared to using the Flask framework alone.
\end{itemize}

The rest of this paper is structured as follows.
\Cref{sec:prelim} introduces recent algorithms for Non-Intrusive Load Monitoring and presents related works on edge-cloud collaboration. \Cref{sec:framework} provides a detailed description of our proposed edge-cloud framework and the NILM model.
\Cref{sec:implement} describes the implementation details for deploying the proposed framework in real scenarios.
\Cref{sec:result} presents the experimental results, and the conclusion is presented in \Cref{sec:conclu}.

\section{Related Work}\label{sec:prelim}

\subsection{Non-Intrusive Load Monitoring}

Intrusive Load Monitoring (ILM) requires a sensor for each monitored device, leading to high costs. In response, Non-Intrusive Load Monitoring (NILM) has been proposed, requiring only the measurement of aggregated data and using decomposition algorithms to estimate individual device consumption. NILM technology was originally introduced to minimize the number of installed electricity meters, thereby reducing wiring harnesses and increasing retrofit capacity. Over the past decades, NILM has found applications in various fields, including public administration and energy management, such as optimizing load schedules in smart grids and improving customer satisfaction\cite{b1}. In private sectors, NILM technology is used for fault detection and diagnosis in both industrial and residential settings\cite{b2}, and it can also evaluate socioeconomic information and consumer behavioral patterns~\cite{b5}.

Since appliances usually exhibit different electrical and operational characteristics, non-intrusive load monitoring techniques can obtain power aggregate signals and then estimate the energy consumption of individual appliances using decomposition algorithms. NILM methods can be divided into three main categories: Machine Learning (ML), Pattern Matching (PM), and Single-Channel Source Separation (SS)\cite{b6}. With the continuous advancement of artificial intelligence, most mainstream NILM techniques today are based on machine learning. The core idea is to extract features from the electrical data and use them to train machine learning algorithms.

At the same time, the development of deep learning and big data has led to a significant increase in data-driven methods using large-scale datasets. Models such as Convolutional Neural Networks (CNNs), Recurrent Neural Networks (RNNs)\cite{b7}, and Long Short-Term Memory networks (LSTM) \cite{b8} have been used to solve NILM problems. Some studies have focused on models such as Generative Adversarial Networks (GAN) and Transformer\cite{b9} to integrate self-attention mechanisms to further improve the performance of decomposition algorithms. At present, the most mainstream models are Seq2Seq\cite{b10}, Seq2Point\cite{b11}, and their variants. Too many parameters will lead to long inference time and greatly reduce practicability. Similarly, some research focuses on the design of lightweight networks and the application of methods such as federated learning to realize parameter sharing.

\subsection{Edge-Cloud Collaboration}
With the advancement of deep learning and edge hardware, there is a growing trend to integrate the two paradigms of cloud computing and edge computing. Edge-cloud collaboration can fully merge the benefits of each and has been implemented in some scenarios. In the smart grid, technologies such as edge computing and the Internet of Things can provide comprehensive control and monitoring solutions to improve energy efficiency, reliability, and real-time service response time\cite{b12}. In the NILM system, it can enhance overall applicability. For example, it can shift the processing and storage of power data, NILM model decomposition, and other tasks to the cloud, making the system easier to use and maintain\cite{b13}. With the support of the cloud's ample computing power, the resource limitation of the edge side can be overcome, and the accuracy of the NILM algorithm can be enhanced\cite{b14, b17}. The edge side can carry out operations such as data cleaning, preprocessing, and even machine learning model deployment if the computing power allows. This can reduce the amount of data received by the cloud, lessen the load, and reduce the system's cost.

Most existing edge-cloud collaborative architectures applied to NILM are two-tier. Smart meters and similar hardware perform data collection and processing at the edge, while servers handle load decomposition in the cloud. In some implementations, load decomposition is conducted at the edge, with the cloud responsible for data dumping and business services. The proposed structure differs by decoupling the edge-end and providing a more fine-grained division of labor at the edge. Data acquisition can be fully delegated to the CT or the meter, greatly reducing the demands on the original hardware performance in the monitored environment. Furthermore, the additional edge-end system carrier can be chosen according to the specific task, lowering deployment costs while maintaining the architecture's benefits.

\section{Framework of Edge-Cloud Collaboration}\label{sec:framework}
\begin{figure*}[htbp]
    \centering
    \includegraphics[width=0.89\linewidth]{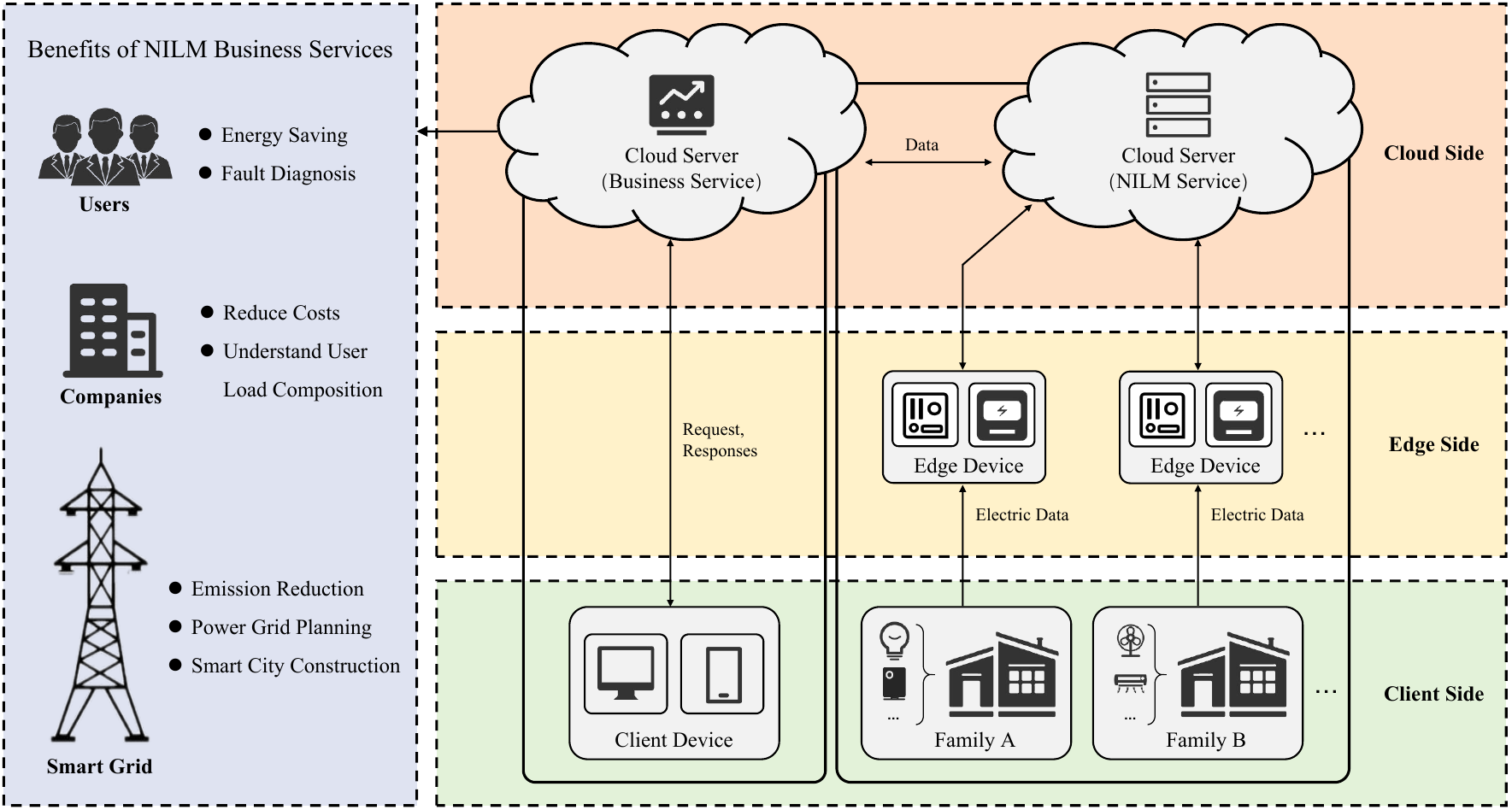}
    \caption{An overview of the three-tier NILM system architecture with edge-cloud collaboration.}
    \label{fig:overview}
    \vspace{-0.3cm}
\end{figure*}
Currently, most NILM systems function exclusively within cloud environments, neglecting the potential benefits of client-side collaboration. This reliance on cloud-based processing leads to substantial consumption of computing resources and significant latency. To address these issues, we propose an innovative edge-cloud collaborative NILM framework. This framework aims to optimize the deployment of NILM models by distributing relatively light computational tasks to edge devices, such as data preprocessing, thereby enhancing efficiency and reducing delays. Furthermore, our study examines the applicability and effectiveness of this hybrid approach in real-world scenarios.

\subsection{Overview}

Most existing edge-cloud architectures adopt a two-tier structure, which leads to high computational costs and compromises on latency. To address this limitation, we propose a novel three-tier architecture comprising a central cloud, edge cloud, and user, where the edge and central clouds synergistically collaborate to deliver services to users. The proposed system architecture is illustrated in Fig.~\ref{fig:overview}. The detailed design of the three components - cloud side, edge side, and client side - is presented below.

\paragraph{Cloud Side} The cloud side consists of a powerful server with substantial computing resources. In the proposed architecture, the NILM server is designed to operate independently of the business server, allowing for greater flexibility and scalability.
This separation is necessary because NILM requires the implementation of a deep learning model, which can be efficiently supported by a microservice architecture. This approach facilitates more nuanced business differentiation, while also providing enhanced maintainability, scalability, and portability.

\paragraph{Edge Side} The edge side comprises two primary components: terminal equipment and a system carrier. Terminal equipment mainly refers to smart meters. Notably, smart meters can be substituted with hardware devices featuring monitoring capabilities, such as smart CT devices. Currently, various brands offer products for power monitoring, including the Sense Energy Monitor, Emporia Vue Smart Home Energy Monitor, TP-Link Kasa Smart Wi-Fi Plug with Energy Monitoring, and Zendure Satellite Monitor CT, among others. The system carrier can be any hardware that supports data preprocessing and model deployment, with its performance contingent upon the system's required functionality. It is worth noting that if the smart meter possesses sufficient performance and scalability, the need for a system carrier may be obviated.

\paragraph{Client Side} The client side of the system comprises the end-users, typically comprising individuals and their families, who reside in households with diverse electrical appliances. These appliances can range in size from small devices like heaters and light bulbs to large equipment such as washing machines and air conditioners. When a user initiates a request, the system's NILM service is triggered, facilitating the analysis of the electrical consumption patterns of these households. This service enables the system to provide personalized insights, recommendations, and feedback to the users, empowering them to optimize their energy usage, reduce their carbon footprint, and improve their overall quality of life.

\subsection{Communication Middleware}

During the implementation of the proposed NILM framework, we encountered two significant challenges in the actual application scenario:

\begin{itemize}
    \item \textbf{Asynchronous processing}: Electrical data is not always delivered to the model immediately after it is sent to the cloud. Although data is transmitted in real-time, batch inference is typically performed after reaching a predetermined threshold to reduce the consumption of computing resources, resulting in an asynchronous processing phenomenon.
    \item \textbf{High concurrent traffic}: NILM systems need to process a large amount of electrical information from different homes simultaneously. When a large number of users request NILM services at the same time, the system can use RabbitMQ as a buffer to ensure that the service runs normally.
\end{itemize}

\begin{figure}[t]
    \centering
    \includegraphics[width=0.93\linewidth]{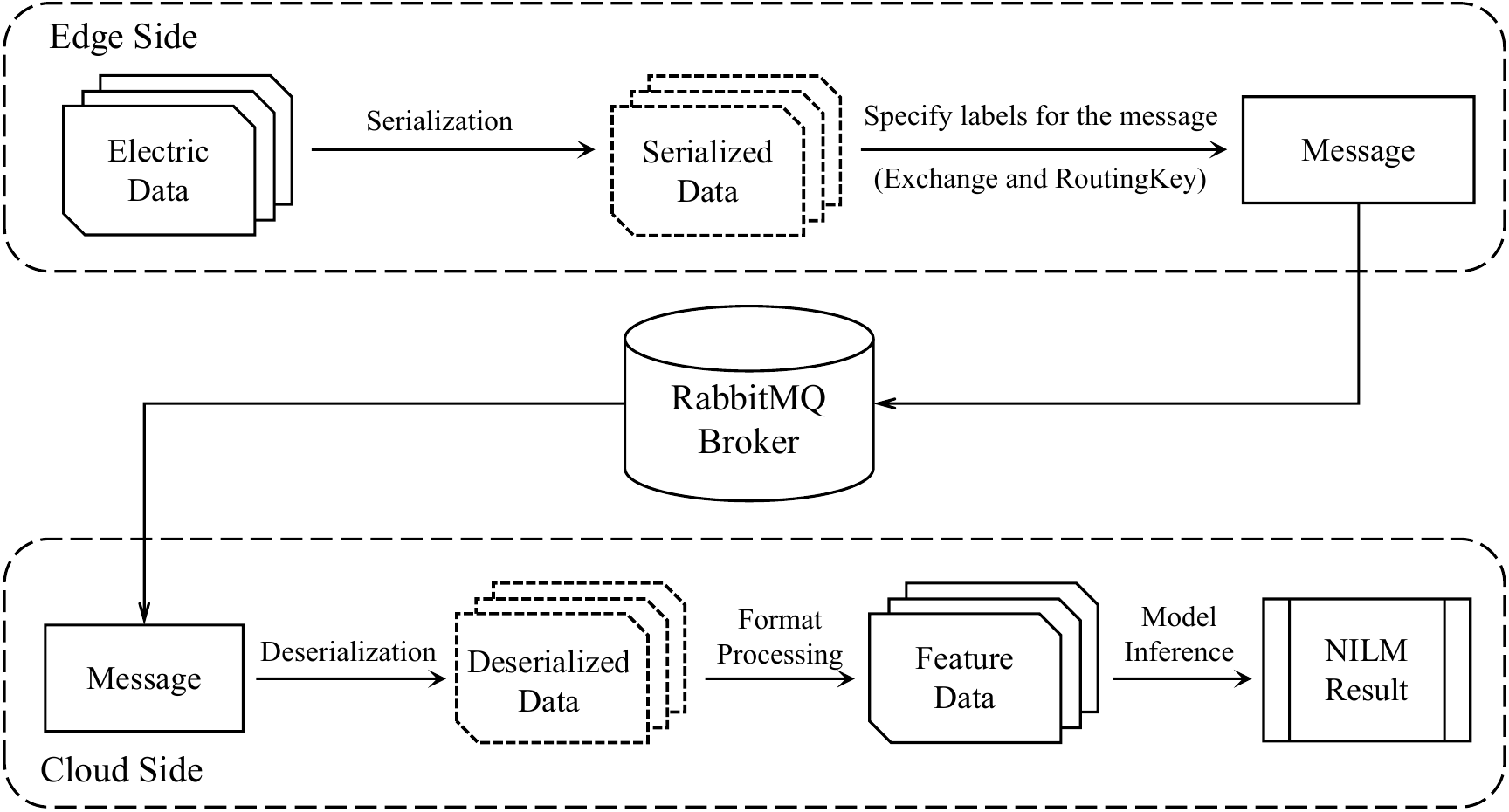}
    \caption{The process of communication between the edge and the cloud through RabbitMQ message queues. The RabbitMQ Broker consists of multiple switches and queues.}
    \label{fig:communi}
    \vspace{-0.5cm}
\end{figure}

Therefore, we propose to leverage the RabbitMQ Message Queue to achieve efficient communication between the cloud and edge. RabbitMQ is an open-source messaging middleware that implements AMQP(Advanced Message Queuing Protocol) in Erlang. It is widely used by various companies because of its reliability, flexible messaging policy, and support for clustering. The communication process of edges and clouds in the system is illustrated in Fig.~\ref{fig:communi}, which highlights the key interactions and data flows between these entities. To begin with, on the edge side, the data producer converts the preprocessed data into a standardized JSON format and sends packets to the RabbitMQ Message Queue in a sequential manner, allowing for efficient batching and handling of large volumes of data. Subsequently, serving as the receiver, the cloud side unpacks the received data, extracts the feature data, and delivers it to the model for further inference, thereby enabling timely and accurate decision-making.
\subsection{NILM Model}
\begin{figure*}[htbp]
    \centering
    \subfigure[Seq2Point Model Architecture.]
    {\label{fig:s2p}\includegraphics[width=0.74\linewidth]{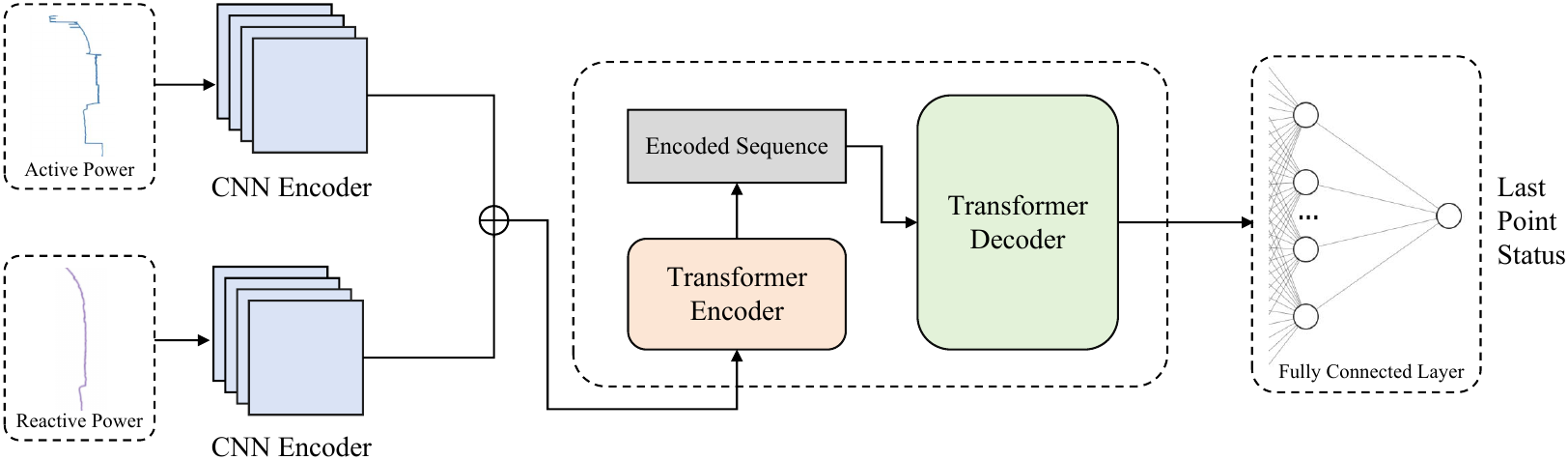}}
    \subfigure[Transformer Module.]{\label{fig:trans}\includegraphics[width=0.20\linewidth]{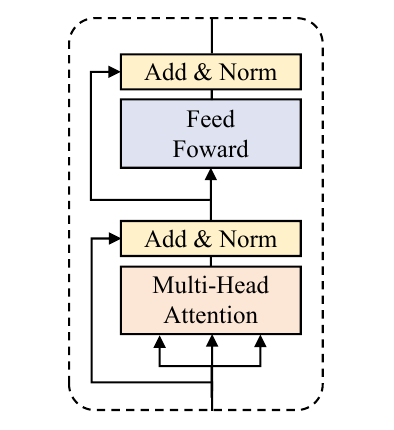}}
    \hfill
    \caption{Deep learning model architecture used in the cloud, where Fig.~\ref{fig:trans} is the structure of Transformer Encoder and Decoder in Fig.~\ref{fig:s2p}. The Multi-Head Attention module in Fig.~\ref{fig:trans} can be superimposed.}
    \vspace{-0.3cm}
\end{figure*}
We propose distinct decomposition models tailored to the cloud and edge environments. Due to its limited computing power, the edge requires a more lightweight model, whereas the cloud's abundant computing resources can support more complex network architectures and larger models, leading to enhanced accuracy in NILM. By leveraging the strengths of both edge and cloud, we can achieve a more precise and efficient NILM service for clients. The details of both edge and cloud models are shown below.
\paragraph{Edge-End Lightweight Model}
For NILM model deployment on the edge, we use the lightweight XGBoost\cite{b18}. More specifically, XGBoost (eXtreme Gradient Boosting) is an implementation of the Gradient Boosting Decision Tree (GBDT) that relies on the gradient boosting algorithm. This ensemble learning technique corrects the errors of previous models by progressively adding new models, each focusing on samples where the previous model's predictions were inaccurate. XGBoost leverages decision trees as a base learner and employs the gradient descent algorithm to minimize the loss function, guiding the model's error correction and updates. Unlike traditional GBDT, XGBoost includes a regularization term in its objective function, helping to control model complexity and prevent overfitting, thereby improving generalization capability. Due to its lightweight nature, hardware devices at the edge, such as smart CT devices, can use XGBoost for NILM prediction. However, for complex decomposition problems, XGBoost alone at the edge is insufficient, necessitating the use of cloud-based NILM models.
\paragraph{Cloud-End Oriented Deep Learning Model}
To ensure decomposition accuracy, the Seq2Point model \cite{b11} is employed on the cloud side. The Seq2Point model architecture is a variant of Seq2Seq, where the midpoint of the output window is represented as a nonlinear regression of the total power window. In other words, the state of the midpoint is related to the electrical information before and after it. The input to the Seq2Point model is the total power \(y_t\), and the output is the predicted state of the midpoint \(X_t\) of the output window. The loss function of the neural network is\cite{b11}:
\begin{equation}
L_p=\sum\limits_{t=1}^{T-W+1}\log p(x_\tau\mid Y_{t:t+W-1},\theta_p)\label{e2}
\end{equation}
where \(\theta_p\) is the parameter of network \(f_p\). 

Based on the Seq2Point model, the input is processed through a sliding window during training and inference, implemented using a queue data structure in actual deployment. After data processing, we develop a neural network architecture that includes convolutional encoders and a Transformer model to predict the state of each household appliance. The structure of the neural architecture is shown in Fig.~\ref{fig:s2p}. The model's input consists of a multidimensional time series, with active and reactive power as features. After processing the sequence features, the multidimensional convolutional encoder concatenates the results and passes them to the subsequent Transformer model.

To better capture information in long time series, we also use Transformer Encoder and Decoder \cite{b9}. The Transformer architecture, based on the self-attention mechanism, addresses the problem of gradient explosion in traditional models such as LSTM \cite{b8} and RNN \cite{b7} when dealing with long sequences. Its key feature is the complete abandonment of the traditional recurrent neural network architecture in favor of a self-attention layer, allowing the model to parallelize computations more efficiently when processing sequential data and better capture long-distance dependencies. The final result represents the state of each appliance at the last point of the sequence.
\section{Implementation}\label{sec:implement}

\subsection{Data Collection and Preprocessing}

Nowadays, most NILM methods are data-driven, and sufficient, high-quality data can greatly improve the efficiency and accuracy of NILM models. Household data is primarily collected through monitoring equipment and system carriers at the edge. The monitoring device can detect current, voltage, active power, reactive power, and other information of the corresponding device through its clamp structure (or a similar method). To meet the data requirements of the NILM model, the monitoring device must achieve a certain sampling rate. These collected data are then sent to the system carrier via a fixed communication protocol. The system carrier is responsible for cleaning and formatting the raw data and sending the processed data to the NILM cloud server through the RabbitMQ queue, where the data is also stored. This process reduces the transmission and storage resource consumption caused by dirty data and decreases the data-processing load on the cloud side. Additionally, the configuration of the edge-end system carrier can be adjusted according to the task. If only data preprocessing is required, lower-cost hardware such as an MCU can be used to reduce costs.

\subsection{NILM Model Development and Deployment}

Many existing deep learning models can be applied to NILM, though some additional data processing may be required for different model architectures. If the decomposition model is based on supervised learning, it is necessary to manually label the switching states of the appliances in the data. There are various ways to obtain appliance labels; for instance, a separate monitoring device can be installed for each appliance to collect labeled data.

The general process of leveraging deep learning models for NILM tasks involves model design, training, and evaluation. After developing the deep learning models, the structure and parameters need to be saved, either in the original training format or converted to a standard format (e.g., ONNX, PMML). The standard format allows the deployment environment to be isolated from the training environment. Using the raw format eliminates the need to revalidate the inference but complicates the deployment environment. This deployment challenge is addressed in this paper by using Docker. 

\subsection{System Building and Testing}

NILM services can be categorized into two types: online prediction and offline prediction. Online prediction is typically a real-time service, requiring the system to provide an API for client calls, with the REST interface being the most popular. Offline prediction involves setting a fixed task on the server, reading data from the database for prediction, and then storing the results back in the database for subsequent access. When constructing the system, it is essential to determine the service type based on different business requirements and then design the specific system architecture accordingly.

\begin{figure}[htbp]
    \centering
    \includegraphics[width=1\linewidth]{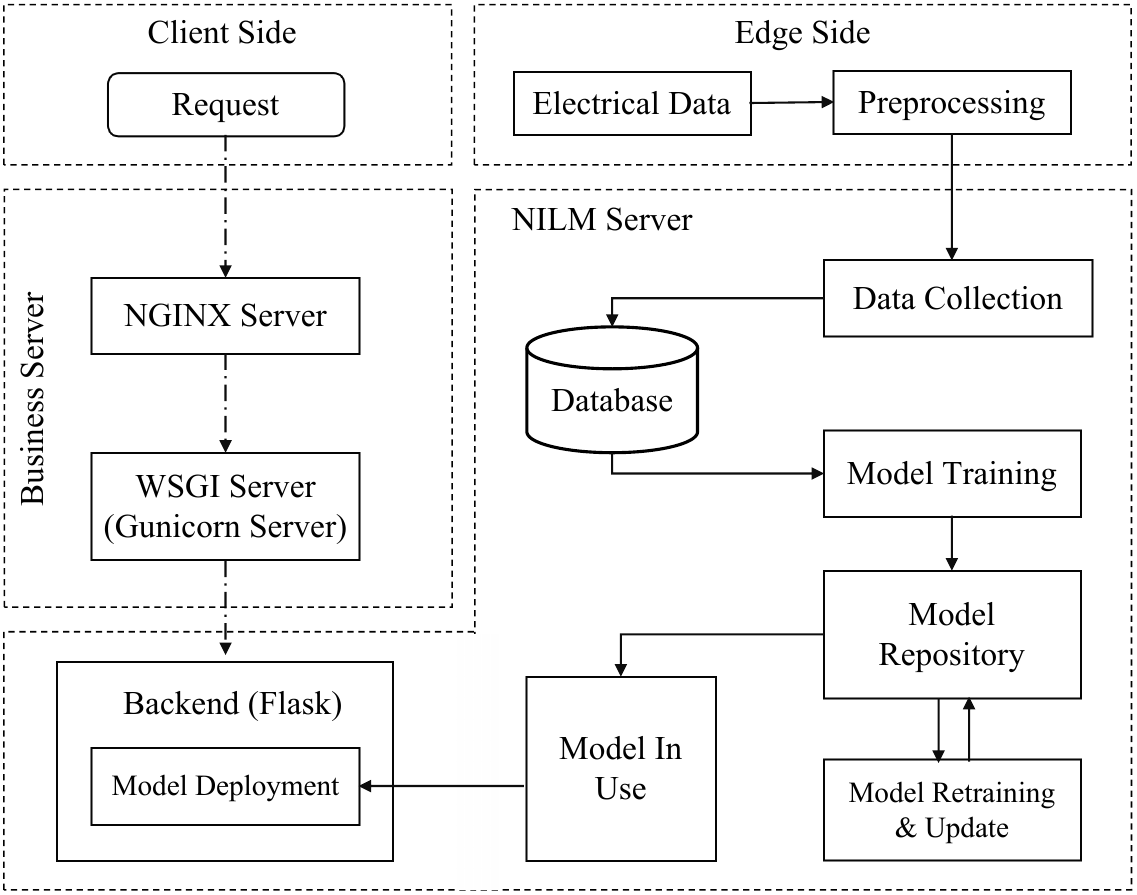}
    \caption{The detailed deployment scheme we adopt in the cloud and the process of responding to client requests.}
    \label{fig:structure_cloud}
\end{figure}

To facilitate model deployment, the proposed system is primarily developed in Python and offers online prediction services, as household data are collected in real time. The overall deployment scheme at cloud is shown in Fig.~\ref{fig:structure_cloud}, which comprises the following components:
\paragraph{Backend and Interface}The NILM model needs to provide uWSGI services and REST APIs using a web backend framework. Common frameworks include Django, Flask, and Falcon, with Flask being particularly convenient for encapsulating models as microservices. However, in production environments, frameworks like Flask and FastAPI often do not meet performance requirements on their own. Therefore, components such as Gunicorn and NGINX need to be integrated to provide efficient concurrency, load balancing, and other functionalities.
\paragraph{NGINX Server}
Nginx is an asynchronous framework web server that serves as a reverse proxy, load balancer, and HTTP cache. It features low memory usage, quick startup, and strong concurrency.
\paragraph{WSGI Server(Gunicorn Server)}
Flask's native server is not suitable for production environments, and using only NGINX as a reverse proxy can lead to unresponsiveness. Therefore, we propose incorporating the Gunicorn server to address this performance issue.
\paragraph{Database}The type of database is determined based on the overall data level of the system, read/write ratio, and other requirements. Commonly used databases include MySQL, Oracle, etc. In addition, Redis can be introduced to improve the caching performance of the system.

\section{Experiments}\label{sec:result}
\subsection{Experimental System and Settings}\label{sec:setting}
We propose a novel three-tier NILM framework based on edge-cloud collaboration to deliver fast and accurate NILM services for clients. To validate the method's effectiveness in real-world settings, we create a home-like environment through laboratory simulations. This environment includes typical household electrical appliances, and we simulate their operational states based on common household behaviors. The collected electrical data will be used for training the NILM model and testing the edge-cloud collaborative architecture.

\paragraph{Datasets}The datasets used to train and assess the model are gathered in a laboratory environment. The original data, sampled every 2 seconds, is cleaned and relabeled before being used to develop the relevant machine learning models. These datasets include variables such as timestamp, mains voltage, frequency, current transformer (CT) probe readings, active power, mains current, reactive power, apparent power, power factor, and load labels. The datasets feature six types of electrical appliances: air purifier, heater, light bulb, fan, air compressor, and air conditioner. 

In addition to appliance type, we also consider different appliance levels as decomposition targets. Some appliances are represented by models from various manufacturers. Due to changes in the experimental environment, two distinct datasets are used for training and evaluating the edge-end and cloud-end models. Given the sparsity of heater data in cloud datasets, we consider the two levels as one.

\paragraph{Edge and Cloud System Carrier}In our proposed three-tier NILM system, the monitor CT functions as an edge monitoring device, enabling real-time monitoring of the power consumption of bus and terminal equipment within the electrical system. The edge system is supported by a Raspberry Pi 5, which has a 64-bit quad-core Arm Cortex-A76 processor and 8 GB of RAM. For the cloud system, we use a server equipped with an Intel i7-10875H processor, 16 GB of RAM, and an RTX 2060 GPU.
\paragraph{Evaluation Metrics}We employ \(\mathit{Accuracy}\), \(\mathit{Recall}\), \(\mathit{Precision}\) and \( \mathit{F_1\text{-}Score}\) to evaluate the performance of NILM models. 

The \( \mathit{F_1\text{-}Score} \) for each appliance is calculated by
\begin{equation}
\begin{aligned}
F_1&=\frac{2\times Precision\times Recall}{Precision+Recall}\\Precision&=\frac{TP}{TP+FP}, Recall=\frac{TP}{TP+FN}
\end{aligned}
\end{equation}

The \(\mathit{Accuracy}\) for each appliance is calculated by
\begin{equation}
Accuracy=\frac{TP+TN}{TP+TN+FP+FN}
\end{equation}
where \(\mathit{TP}\) is the number of true positives, meaning the appliance is in the ON state and correctly identified; \(\mathit{TN}\) is the number of true negatives, meaning the appliance is in the OFF state and correctly identified; \(\mathit{FP}\) is the number of false positives, meaning the appliance is actually in the OFF state but identified as ON; \(\mathit{FN}\) is the number of false negatives, meaning the appliance is actually in the ON state but identified as OFF.

The NILM system service performance is evaluated using four metrics: Average Response Time, Median Response Time, 90\% Response Time, Max Response Time, and Throughput.
\subsection{Evaluations of NILM Model}

To demonstrate the effectiveness of the NILM model, we use the collected laboratory datasets to compare the performance of edge-end and cloud-end NILM models, specifically XGBoost and Seq2Point. The evaluation results for the edge-side NILM model are shown in Table~\ref{tab:xgbmodel}. These results indicate a relatively high accuracy (~92.6\%) yet a low F1-score (74.1\%) on average. This discrepancy may be due to the data sparsity of some appliances, such as different levels of the fan, and the limited expressiveness of the lightweight XGBoost model. In contrast, the decomposition results of the cloud-end model, shown in Table~\ref{tab:s2pmodel}, demonstrate both satisfactory accuracy and F1-score, highlighting the effectiveness of our proposed cloud-end model structure that combines CNN and Transformer. However, the Seq2Point model on the cloud end requires significantly more computational resources than the edge-end model. Therefore, by combining cloud and edge models, we can provide accurate and cost-effective services to clients.
\begin{figure*}[htbp]
    \centering
    \subfigure[Flask-only.]{\label{fig:res1}\includegraphics[width=0.3\linewidth]{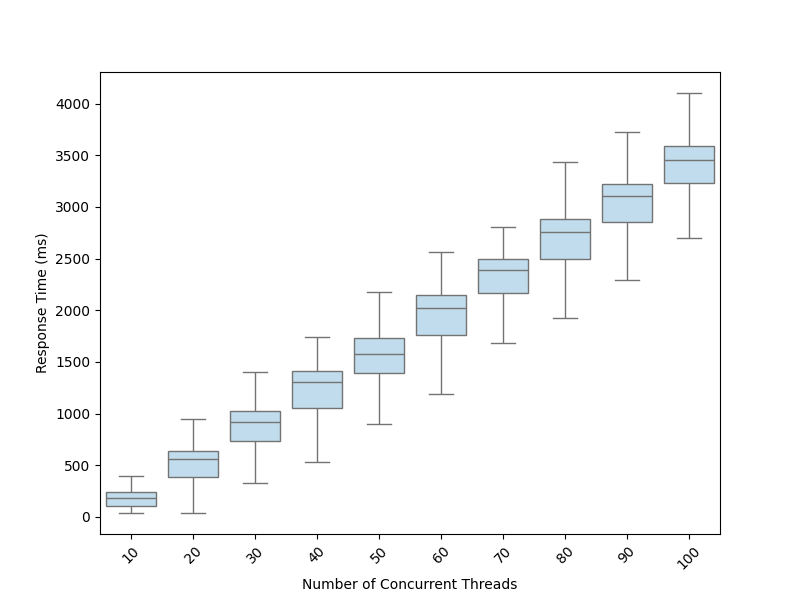}}
    \subfigure[Flask, NGINX and 2 Workers Gunicorn.]{\label{fig:res2}\includegraphics[width=0.3\linewidth]{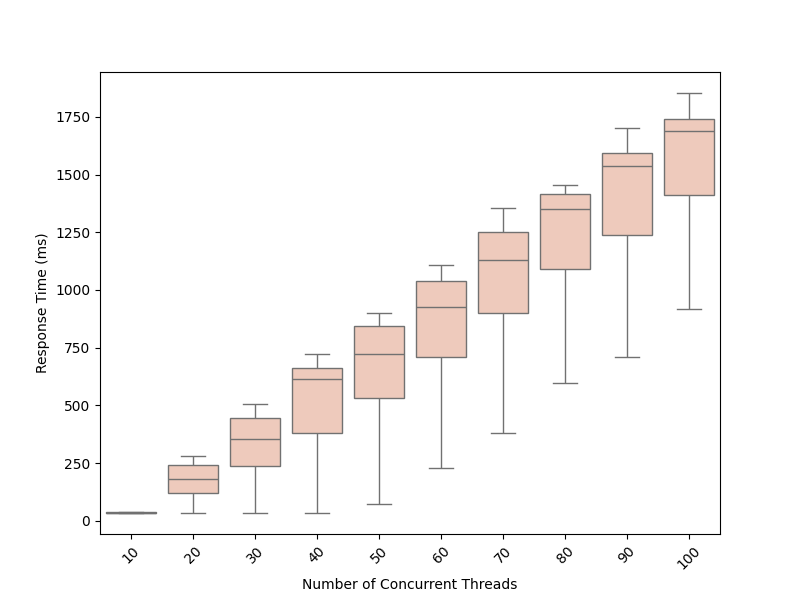}}
    \subfigure[Flask, NGINX and 4 Workers Gunicorn.]{\label{fig:res3}\includegraphics[width=0.3\linewidth]{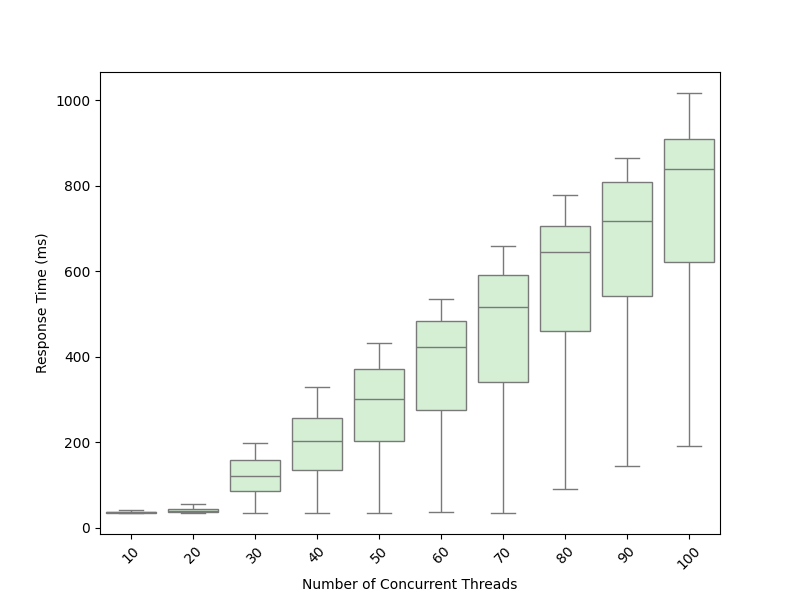}}
    \caption{Comparison of the respective response times of the concurrent architecture and the infrastructure under different numbers of concurrent threads.}
    \label{fig:result}
    \vspace{-0.3cm}
\end{figure*}
\begin{table}[bhtp] 
\vspace{-0.2cm}
\caption{Evaluation results of edge-end NILM model (XGBoost)}
\begin{center}
\begin{tabular}{ccccc}
\toprule Appliance & Accuracy & Recall & Precision & F1-Score \\
\midrule 
    heater\_1 & 0.9946 & 0.9835 & 0.9952 & 0.9893 \\
    heater\_2 & 0.9942 & 0.9693 & 0.9879 & 0.9781 \\
    air\_purifier\_A & 0.8136 & 0.6914 & 0.7084 & 0.6773 \\
    air\_purifier\_B & 0.8454 & 0.6933 & 0.7275 & 0.6974 \\
    fan\_1 & 0.8992 & 0.6101 & 0.5444 & 0.5570 \\
    fan\_2 & 0.8255 & 0.5589 & 0.5085 & 0.4548 \\
    fan\_3 & 0.9251 & 0.6378 & 0.7656 & 0.6624 \\
    fan\_4 & 0.9181 & 0.3870 & 0.5426 & 0.4425 \\
    light\_bulb$*$1 & 0.9795 & 0.7833 & 0.9557 & 0.8363 \\
    light\_bulb$*$2 & 0.9873 & 0.9156 & 0.9206 & 0.9097 \\
    air\_compressor & 0.9993 & 0.9649 & 0.9512 & 0.9506 \\
\midrule
    Average & 0.9256 & 0.7450 & 0.7825 & 0.7414 \\
\bottomrule
\vspace{-0.2cm}
\end{tabular}
\end{center}
\label{tab:xgbmodel}
\end{table}
\begin{table}[bhtp] 
\vspace{-0.2cm}
\caption{Evaluation results of cloud-end NILM model (Seq2Point)}
\begin{center}
\begin{tabular}{ccccc}
\toprule Appliance & Accuracy & Recall & Precision & F1-Score \\
\midrule 
    air\_purifier\_A & 0.9134 & 0.9944 & 0.8315 & 0.9057 \\
    heater\_total & 0.9753 & 0.9992 & 0.9348 & 0.9625 \\
    light\_bulb\_1 & 0.9882 & 0.9899 & 0.9315 & 0.9598 \\
    light\_bulb\_2 & 0.9887 & 0.9981 & 0.9427 & 0.9615 \\
    air\_compressor & 0.9990 & 0.9218 & 0.8462 & 0.8824 \\
    air\_conditioner & 0.9836 & 0.9984 & 0.9516 & 0.9000 \\
\midrule
    Average & 0.9747 & 0.9796 & 0.9064 & 0.9411 \\
\bottomrule
\end{tabular}
\end{center}
\vspace{-0.4cm}
\label{tab:s2pmodel}
\end{table}
\subsection{Evaluation of Edge-Cloud Collaboration Framework}

\paragraph{Impact of Edge-Side Data Preprocessing}
To demonstrate the impact of data preprocessing at the edge, we extracted one hour of data daily for a week. The edge server successfully filters out 230 dirty data points from 12,758 total data points, reducing the transmission cost by approximately 1.8\%. These data points may have field defaults or negative values due to firmware versions or environmental impacts.

\paragraph{Impact of Model Deployment On Response Time}

We deploy the NILM model exclusively on the edge and exclusively on the cloud to compare the response times of the services provided by these two approaches. We conduct concurrent tests on the microservice interface of model inference to verify the effectiveness of this architecture. We test scenarios with 1, 3, 5, 10, 30, 50, and 100 concurrent threads, each with an access interval of 2 seconds, and repeat each test 10 times. The response times for deploying the model solely on the edge are shown in Table~\ref{tab:rt2}, while the response times for deploying the model solely on the cloud are shown in Table~\ref{tab:rt1}. By comparing these tables, we find that deploying the model on the edge can significantly reduce service response time, especially with many concurrent threads. Notably, with 100 concurrent threads, the response time of the edge-deployed model is only 16.7\% of the response time of the cloud-deployed model. This is because after the model inference is placed on the edge side, the cloud only needs to call the resulting data.

\paragraph{Impact of Load Balancing Module}
To improve the system's concurrent performance and achieve load balancing, we adopt the deployment scheme shown in Fig.~\ref{fig:structure_cloud}. We test the interface in the range of 0-100 concurrent threads, each with an access interval of 2 seconds, and repeat each test 10 times. The comparison results of different deployment schemes are shown in Fig.~\ref{fig:result}, from which we make the following observations: 1) The average response time of deploying with NGINX and Gunicorn with the Flask framework is much lower than deploying with the Flask framework alone, highlighting the necessity of incorporating the NGINX and Gunicorn module. 2) When using Gunicorn as a WSGI server, the number of workers affects performance; more workers generally result in shorter response times. For example, the response time is shortest when there are 4 workers. Therefore, in a real deployment, the number of workers should be chosen based on the server configuration.

It should be noted that all our previous experiments were conducted with a response error rate of 0. However, the NILM service interface does experience response errors when the number of concurrent threads becomes too high. With the configuration described in \Cref{sec:setting}, we find that the basic Flask deployment starts failing frequently after 200 concurrent threads, whereas the NGINX and Gunicorn deployment with the Flask framework only begins to fail after 400 concurrent threads. Overall, this proposed deployment scheme significantly increases the threshold and enhances the reliability of the service.
\begin{table*}
\caption{Response Time When NILM Model Solely Deployed on the Edge Side}
\begin{center}
\begin{tabular}{cccccc}
\toprule Concurrency & Average(ms) & Median(ms) & 90\% Line(ms) & Max(ms) & Throughput(TPS/s)\\
\midrule 
    1 & 8 & 8 & 9 & 60 & 112.1 \\
    3 & 11 & 9 & 16 & 30 & 123.8 \\
    5 & 27 & 31 & 39 & 81 & 114.3 \\
    10 & 72 & 76 & 102 & 172 & 109.6 \\
    30 & 267 & 273 & 383 & 964 & 103.7 \\
    50 & 461 & 470 & 763 & 2421 & 102.9 \\
    100 & 922 & 965 & 1738 & 4327 & 100.4 \\
\bottomrule
\end{tabular}
\end{center}
\label{tab:rt2}
\end{table*}
\begin{table*}
\vspace{-0.2cm}
\caption{Response Time When NILM Model Solely Deployed on the Cloud Side}
\begin{center}
\begin{tabular}{cccccc}
\toprule Concurrency & Average(ms) & Median(ms) & 90\% Line(ms) & Max(ms) & Throughput(TPS/s)\\
\midrule 
    1 & 59 & 59 & 61 & 81 & 16.8 \\
    3 & 157 & 167 & 210 & 272 & 17.5 \\
    5 & 269 & 278 & 333 & 546 & 17.6 \\
    10 & 551 & 558 & 663 & 875 & 17.6 \\
    30 & 1666 & 1691 & 1932 & 2395 & 17.5 \\
    50 & 2816 & 2846 & 3318 & 3918 & 17.3 \\
    100 & 5531 & 5666 & 6103 & 6514 & 17.3 \\
\bottomrule
\end{tabular}
\end{center}
\label{tab:rt1}
\vspace{-0.3cm}
\end{table*}


\section{Conclusion}\label{sec:conclu}
In this paper, we propose a three-tier framework as well as collaborative edge-cloud schemes to address the challenges in applying NILM in real-world deployment. Our solution reduces service response time, alleviates cloud workload, and enhances data security. Unlike previous works that focus primarily on the algorithm level, neglecting real-world deployment details, we systematically design the NILM serving architecture to achieve load balancing and low-latency service. We implement XGBoost and Seq2Point models for load decomposition on the edge and the cloud, respectively. Experimental results demonstrate that edge deployment can reduce service response time and communication overhead, while cloud deployment offers more accurate load decomposition and facilitates management. Furthermore, by integrating Gunicorn and NGINX, we achieve load balancing and address cloud-side response errors during high concurrency. By paving the way toward real-world NILM deployment, this work provides a foundational framework for future in-depth NILM research, allowing for the exploration of personalized solutions for different households and the iterative update of models.


\begin{thebibliography}{00}

\bibitem{b1} H. Cimen, N. Cetinkaya, J. C. Vasquez, and J. M. Guerrero, “A Microgrid Energy Management System Based on Non-Intrusive Load Monitoring via Multitask Learning,” IEEE Trans. Smart Grid, vol. 12, no. 2, pp. 977–987, Mar. 2021, doi: 10.1109/TSG.2020.3027491.
\bibitem{b2} A. Aboulian et al., “NILM Dashboard: A Power System Monitor for Electromechanical Equipment Diagnostics,” IEEE Trans. Ind. Inf., vol. 15, no. 3, pp. 1405–1414, Mar. 2019, doi: 10.1109/TII.2018.2843770.
\bibitem{b3} H. Hong, J. Huan, X. Pan, Y. Sui, X. Zhang, and X. Jiang, “Energy Usage Identification Method of Integrated Energy System Based on Edge-Cloud Coordination,” in 2021 IEEE Asia-Pacific Conference on Image Processing, Electronics and Computers (IPEC), Dalian, China: IEEE, Apr. 2021, pp. 258–262. doi: 10.1109/IPEC51340.2021.9421217.
\bibitem{b4} N. Hudson, M. J. Hossain, M. Hosseinzadeh, H. Khamfroush, M. Rahnamay-Naeini, and N. Ghani, “A Framework for Edge Intelligent Smart Distribution Grids via Federated Learning,” in 2021 International Conference on Computer Communications and Networks (ICCCN), Athens, Greece: IEEE, Jul. 2021, pp. 1–9. doi: 10.1109/ICCCN52240.2021.9522360.
\bibitem{b5} P. A. Schirmer and I. Mporas, “On the non-intrusive extraction of residents’ privacy- and security-sensitive information from energy smart meters,” Neural Computing and Applications, vol. 35, no. 1, pp. 119–132, Jan. 2023, doi: 10.1007/s00521-020-05608-w.
\bibitem{b6} P. A. Schirmer and I. Mporas, “Non-Intrusive Load Monitoring: A Review,” IEEE Trans. Smart Grid, vol. 14, no. 1, pp. 769–784, Jan. 2023, doi: 10.1109/TSG.2022.3189598.
\bibitem{b7} W. Zaremba, I. Sutskever, and O. Vinyals, “Recurrent Neural Network Regularization.,” arXiv: Neural and Evolutionary Computing, vol. abs/1409.2329, 2014.
\bibitem{b8} S. Hochreiter and J. Schmidhuber, “Long Short-Term Memory,” Neural Computation, vol. 9, no. 8, pp. 1735–1780, 1997.
\bibitem{b9} A. Vaswani et al., “Attention Is All You Need.” arXiv, Dec. 05, 2017. Accessed: May 31, 2023. [Online]. Available: http://arxiv.org/abs/1706.03762
\bibitem{b10} I. Sutskever, O. Vinyals, and Q. V. Le, “Sequence to Sequence Learning with Neural Networks.,” Advances in neural information processing systems, vol. 27, 2014.
\bibitem{b11} C. Zhang, M. Zhong, Z. Wang, N. Goddard, and C. Sutton, “Sequence-to-point learning with neural networks for non-intrusive load monitoring,” AAAI, vol. 32, no. 1, Apr. 2018, doi: 10.1609/aaai.v32i1.11873.
\bibitem{b12} Q. N. Minh, V.-H. Nguyen, V. K. Quy, L. A. Ngoc, A. Chehri, and G. Jeon, “Edge computing for IoT-enabled smart grid: The future of energy,” Energies, vol. 15, no. 17, p. 6140, Aug. 2022, doi: 10.3390/en15176140.
\bibitem{b13} O. Boiko, A. Komin, R. Malekian, and P. Davidsson, “Edge-Cloud Architectures for Hybrid Energy Management Systems: A Comprehensive Review,” IEEE Sensors J., vol. 24, no. 10, pp. 15748–15772, May 2024, doi: 10.1109/JSEN.2024.3382390.
\bibitem{b14} J. Yao et al., “Edge-Cloud Polarization and Collaboration: A Comprehensive Survey for AI,” IEEE Trans. Knowl. Data Eng., pp. 1–1, 2022, doi: 10.1109/TKDE.2022.3178211.
\bibitem{b15} N. Hudson, M. J. Hossain, M. Hosseinzadeh, H. Khamfroush, M. Rahnamay-Naeini, and N. Ghani, “A Framework for Edge Intelligent Smart Distribution Grids via Federated Learning,” in 2021 International Conference on Computer Communications and Networks (ICCCN), Athens, Greece: IEEE, Jul. 2021, pp. 1–9. doi: 10.1109/ICCCN52240.2021.9522360.
\bibitem{b16} M. Viggiato, R. Terra, H. Rocha, M. T. Valente, and E. Figueiredo, “Microservices in Practice: A Survey Study.” arXiv, Aug. 14, 2018. Accessed: Jul. 17, 2024. [Online]. Available: http://arxiv.org/abs/1808.04836
\bibitem{b17} Y. Zhang et al., “FedNILM: Applying Federated Learning to NILM Applications at the Edge,” IEEE Transactions on Green Communications and Networking, vol. 7, no. 2, pp. 857–868, Jun. 2023, doi: 10.1109/TGCN.2022.3167392.
\bibitem{b18} T. Chen and C. Guestrin, “XGBoost: A Scalable Tree Boosting System,” in Proceedings of the 22nd ACM SIGKDD International Conference on Knowledge Discovery and Data Mining, Aug. 2016, pp. 785–794. doi: 10.1145/2939672.2939785.
\end{thebibliography}
\end{document}